\documentclass[%
reprint,
superscriptaddress,
amsmath,amssymb,
]{revtex4-1}

\usepackage{color}
\usepackage{graphicx}% Include figure files
\usepackage{dcolumn}% Align table columns on decimal point
\usepackage{bm}% bold math
\begin {document}
%section {title}
%\preprint{APS/123-QED}

\title{Homogeneous Connectivity of Potential Energy Network in a Solidlike State of Water Cluster}% Force line breaks with \\
%\thanks{A footnote to the article title}%

\author{Takuma Akimoto}
\email{akimoto@z8.keio.jp}
\author{Toshihiro Kaneko}
\author{Kenji Yasuoka}
\affiliation{%
  Department of Mechanical Engineering, Keio University, Yokohama, 223-8522, Japan
}%

\author{Xiao Cheng Zeng}
\affiliation{
Department of Chemistry, University of Nebraska-Lincoln, Lincoln, Nebraska 68588, United States
}

%\collaboration{MUSO Collaboration}%\noaffiliation

\date{\today}% It is always \today, today,
%  but any date may be explicitly specified

\begin{abstract}
%Potential energy surface plays a crucial role in elucidating structures and dynamics in supercooled liquids, the glass transition, and clusters. 
A novel route to the exponential trapping-time distribution within a solidlike state in water clusters is described. 
We propose a simple homogeneous network (SHN) model to investigate dynamics on the potential energy networks of water clusters. 
In this model, it is shown that the trapping-time distribution in a solidlike state follows the exponential distribution, whereas  
the trapping-time distribution in local potential minima within the solidlike state is not exponential. To confirm 
the exponential trapping-time distribution in a solidlike state,
we investigate water clusters, $($H${}_2$O$)_6$ and $($H${}_2$O$)_{12}$,
by molecular dynamics simulations. 
%for several temperatures. 
These clusters change dynamically from solidlike to liquidlike state and vice versa. 
%We consider trapping time $\tau_\alpha$ in a solidlike state and interevent time $\tau_\beta$ of large fluctuations in potential energy within the solidlike state.
%Analyzing the probability density functions (PDFs) of two trapping times, 
We find that the probability density functions of trapping times in a solidlike state %$\tau_\alpha$ 
are described by the exponential distribution whereas those of interevent times of large fluctuations in potential energy within the solidlike state
%$\tau_\beta$  
follow the Weibull distributions. 
The results provide a clear evidence that transition dynamics between solidlike and liquidlike states in water clusters are well described by 
the SHN model, suggesting that the exponential trapping-time distribution within a solidlike state originates from the 
homogeneous connectivity in the potential energy network. 
%Finally, we discuss a generalization of the SHN model to investigate an origin of a non-exponential trapping-time distribution in small clusters. 
\end{abstract}

% PACS, the Physics and Astronomy
% Classification Scheme.
%\keywords{Suggested keywords}%Use showkeys class option if keyword
%display desired
\maketitle

%\tableofcontents

\section{Introduction}

In the theory of statistical mechanics, the system size, such as the number of particles and volume, generally goes to infinity. 
This idealization is not  ideology because physical quantities in finite size systems close to those in bulk as the system size 
becomes large. However, a situation is completely different when we consider a small system such as atomic clusters. 
For example, the melting points of atomic clusters depend irregularly on the system size \cite{Schmidt1998, Kaneko2011}, 
and configurations of molecules of a cluster change drastically with time \cite{Wales1994, Nishio2009}. 
Little is known about statistical mechanics of small size systems, {\it i.e.}, how to replace  
dynamics to a stochastic description.

Energy landscape is one of the most useful descriptions to elucidate structures and dynamics in supercooled liquids, the glass transition, clusters,  
and proteins \cite{Goldstein1969, Stillinger1995, Debenedetti2001, Doliwa2003, Heuer2008, Ball1996, Wales2004}. 
It is widely believed that motions of phase points on the potential energy surface (PES) can be represented by a stochastic description  
if the phase points are coarse-grained suitably \cite{Odagaki1990, Monthus1996}. 
A configuration of molecules will drastically change when the phase point escapes from a deep valley in the PES (metabasin).
 This process is called an {\it $\alpha$-process} whereas escapes from local 
potential minima within a metabasin 
are called a {\it $\beta$-process} \cite{Stillinger1995}.  In other words, the phase point in the PES will be trapped within a metabasin for long 
times while small transitions between local potential minima within the metabasin occur \cite{Buchner2000}. 
It will physically be possible to found statistical mechanics of small clusters using such concepts because the number of potential 
minima increases exponentially with the system size. 

For small systems of supercooled liquids, 
$\alpha$ and $\beta$ processes are clearly observed in molecular dynamics (MD) simulations \cite{Denny2003}. 
Moreover, statistical properties of hopping times from metabasins 
are characterized by potential energy barriers in the PES. The Arrhenius law tells us that
 the mean trapping time (the mean time to escape from a potential minimum) 
 is proportional to $\exp(\Delta E/k_BT)$, where $\Delta E$ is a height of a potential energy barrier, 
 $k_B$ is the Boltzmann constant, and $T$ is the temperature. 
Based on the Arrhenius law, 
Gaussian trap model, where barrier heights are distributed according to a Gaussian, provides a good stochastic description 
of dynamics in  small systems of supercooled liquids \cite{Denny2003}. 

PES's are highly heterogeneous as well as multi-dimensional. 
A network of PES is composed of basins of attraction around each potential minimum (nodes) and possible 
paths between nodes (links). Although the potential energy network (PEN) does not depend on temperature, 
PENs constructed by MD simulations depend on the temperature  because 
the phase point cannot wander the whole PES in a finite time. 
Especially at low temperatures, the phase point will be trapped in a deep potential minimum.
It has been shown that PENs in small Lennard-Jones clusters have a small-world and scale-free 
character \cite{Watts1998, Barabasi1999, Doye2002}, 
where PENs are constructed by an inherent structure network, and thus do not depend on temperature.
%Transitions between potential minima can be described by random walks on PENs. 
If PENs have a small-world character, %a state can change from a solidlike to liquidlike state and vice versa 
a configuration can change drastically
by  a few steps on PENs
 because almost all nodes are connected through a few nodes due to a small-world character \cite{Watts1998, Barabasi1999}.

%Although static properties of PENs have been known, little is known about dynamical behaviors on PENs. 
Disconnectivity graphs are also used to characterize the PES of clusters \cite{Wales1998a,Doye1999, Wales2000}. 
According to the disconnectivity graph of the TIP4P water cluster $($H${}_2$O$)_6$ \cite{Wales2000},  lower potential minima 
(nodes) such as cage, prism, and book can be connected each other if the temperature of the water cluster is sufficiently large. 
  However, how those nodes are connected and which paths are common  
  remain unclear. 
The graphs alone are insufficient to investigate dynamical properties on the PENs  
because dynamical behaviors depend upon how potential minima are connected \cite{Wales2004, Souza2008, Souza2009}.
 In particular, the connectivities between nodes within a metabasin
 and those within the other metabasins affect transition dynamics between metabasins. 
  In binary Lenard-Jones systems \cite{Souza2008, Souza2009}, transition dynamics related to cage-breaking events 
 can be described as correlated random walk. Here, we study transition dynamics between 
 solidlike and liquidlike states which are defined by coarse-grained potential energy (see details in section III). 
 If nodes within a specific metabasin are sparsely connected to those within the other metabasins, 
 the trapping-time distribution within the specific metabasin will not be the exponential distribution, whereas as shown in the next 
 section, homogeneous connectivities can provide the exponential trapping-time distribution.

 In this paper, we propose a simple homogeneous model  of transition dynamics between specific metabasins called 
 a solidlike state  
 and the other metabasins called a liquidlike state.  
In this model, we analytically obtain a useful relation between the mean 
 trapping times in a solidlike state and local potential minima within the solidlike state, and 
 show that trapping-time distribution in a solidlike state follows 
 the exponential distribution. We confirm these statistical properties for trapping times using MD
 simulations of  water clusters, $($H${}_2$O$)_6$ and $($H${}_2$O$)_{12}$. Finally,
 generalizing the simple homogeneous network model to transition 
 dynamics between specific metabasins and the other metabasins, 
 we discuss an origin of a non-exponential trapping-time distribution in water clusters.

\section{ Simple homogeneous model of potential energy network}

\begin{figure}[h]
\includegraphics[width=1.\linewidth, angle=0]{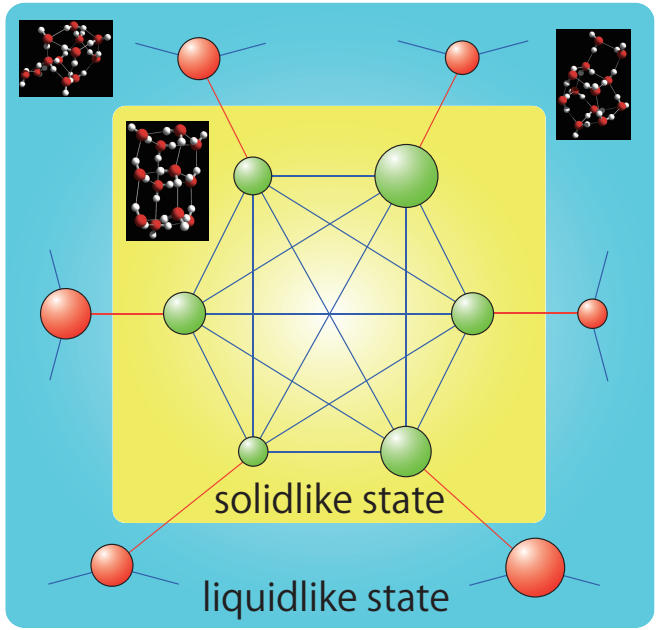}
\caption{ Schematic picture of a simple homogeneous model of a coarse-grained PEN. 
Green and red spheres are potential minima in solidlike and liquidlike states, respectively. 
%We note that values of potential minima are different while we depict the same figure. 
All potential minima in a solidlike state are connected to a potential minimum in a liquidlike state.
Some configurations of solidlike and liquidlike states for $($H${}_2$O$)_{12}$
are shown for reference. 
}
\label{PEN}
\end{figure}

Consider a simple homogeneous network (SHN) model for a PEN (see Fig.~\ref{PEN}). 
In the model, we assume that a coarse-grained phase point on the PEN undergoes a random walk on the 
 PEN with continuous random trapping times ({\it continuous time random walk} \cite{metzler00} on a network). 
Homogeneous means that the trapping-time distribution does not depend on a local potential minimum (node) 
within a specific metabasin, where we call the metabasin a solidlike state and the other metabasins a liquidlike state. 
More precisely, trapping times of all nodes within the solidlike state are independent and identically distributed random variables.  
 Because barrier heights of potential minima are different generally, the above assumption means a coarse graining 
 of nodes  in the solidlike state. 
Moreover, we assume that  the probability that the coarse-grained phase point escapes from the solidlike state  
when it escapes from a node within the solidlike state is always a constant $p$, resulting that the number of trials $k$
to escape from the metabasin (solidlike state) to the other metabasins (liquidlike state) is 
 distributed according to $p_k = p (1-p)^{k-1}$ ({\it geometric distribution}). 
 The constant probability $p$ is related to the connectivity between nodes within the solidlike state and those within the 
 liquidlike state. It is nontrivial but rather surprising that the assumption of the constant probability $p$ is valid 
 in  small clusters. Even when the PEN is a scale-free network, the probability $p_k$ of the number 
 of trials to escape does not follow the exponential distribution if a hub, which has many connectivities to nodes in a liquidlike state, 
 is not included in a solidlike state.
 
The distribution of trapping times $\tau_{\alpha}$ in the metabasin (solidlike state) %, {\it i.e.}, $\alpha$-process, 
 is given by a compound distribution \cite{Feller1971}. 
For example, let $\tau_\beta^1, \tau_\beta^2, \ldots, \tau_\beta^k$ be trapping times at nodes within the 
solidlike state, then trapping time in the solidlike state is given by 
$\tau_{\alpha} = \tau_\beta^1 + \tau_\beta^2 + \ldots + \tau_\beta^k$, where $k$ is the number of steps to escape from the solidlike state
 for the first time. 
The distribution of trapping times $\tau_\alpha$ in the soikdlike state is written as 
\begin{equation}
F(\tau_{\alpha}) = \sum_{k=1}^{\infty} p_k \Pr (\tau_\beta^1 + \tau_\beta^2 + \ldots + \tau_\beta^k < \tau_{\alpha}).
\label{compound}
\end{equation}
%where $p_k$ is the probability of the number of escape from a local minimum until the phase point escapes from the metabasin. 
%{\it Exponential distribution in trapping times within a metabasin}.---
The 
Laplace transform of $F(\tau_\alpha)$, defined by $\tilde{F}(s) = \int_0^{\infty} F(\tau_\alpha) e^{-s\tau_\alpha}d\tau_\alpha$, is given by
\begin{eqnarray}
\tilde{F}(s) &=& p \sum_{k=1}^{\infty} q^{k-1} \{ \tilde{\varphi} (s) \}^{k-1} \tilde{\varphi} (s)/s \\
&=& \frac{1}{s}  \frac{p\tilde{\varphi} (s) }{1 - q \tilde{\varphi} (s)}, 
\end{eqnarray}
where $q=1-p$ and $\tilde{\varphi} (s)$ is the Laplace transform of the PDF $P(\tau_\beta)$ of trapping times at nodes
$\tau_\beta^1, \ldots, \tau_\beta^k$.  We assume that trapping times $\tau_\beta^1, \ldots, \tau_\beta^k$ has a finite mean 
$\langle \tau_\beta \rangle$. 
The mean of $\tau_\alpha$ is given by
\begin{eqnarray}
\langle \tau_{\alpha} \rangle &=& - \left. \frac{d}{ds} \{s \tilde{F}(s)\} \right|_{s=0}\\
&=& -\frac{p\tilde{\varphi}' (0) \{1 - q \tilde{\varphi} (0)\} + p q \tilde{\varphi}(0) \tilde{\varphi}'(0)}{\{1 - q \tilde{\varphi} (0)\}^2}.
\end{eqnarray}
Using $\tilde{\varphi}(0) = 1$ and $\tilde{\varphi}'(0) = -\langle \tau_\beta \rangle$, we have a relation between $\langle \tau_\alpha \rangle$ 
and $\langle \tau_\beta \rangle$: 
\begin{equation}
\langle \tau_{\alpha} \rangle = \frac{\langle \tau_\beta \rangle}{p}.
\label{ave_tau_alpha}
\end{equation}
%where $\langle \tau \rangle$ is the mean of $\tau$.
Using an approximation $\tilde{\varphi}(s) = 1 - \langle \tau_\beta \rangle  s + O(s^2)$ 
for $s\rightarrow 0$, we have 
\begin{eqnarray}
\tilde{F}(s) \cong \frac{1}{s}\frac{1 - \langle \tau_\beta \rangle s}{1+q\langle \tau_\beta \rangle s /p}
\cong \frac{1}{s} \frac{1}{1 + \langle \tau_\beta \rangle s/p}.
\end{eqnarray} 
%(I use $1 - \langle \tau \rangle s \cong (1 + \langle \tau \rangle s)^{-1}$.)} 
The inverse Laplace transform reads 
\begin{equation}
P(\tau_\alpha)= F'(\tau_\alpha)  \sim  \frac{1}{\langle \tau_{\beta} \rangle/p} e^{- \tau_\alpha/ (\langle \tau_{\beta} \rangle/p)}
\quad (\tau_\alpha \rightarrow \infty),
\label{exponential}
\end{equation}
%where $\langle \tau_{\alpha} \rangle= \langle \tau_\beta \rangle/p$.
In the SHN model, the exponential distribution  appears universally in the tail of
 the PDF $P(\tau_\alpha)$  ($\tau_\alpha \rightarrow \infty$). This is similar to the exit-time distribution in  
 a superbasin \cite{Fichthorn2013}. 
We note that the exponential distribution is not originated from the exponential distribution of escape times from a single potential 
valley but from the homogeneous connectivity between the solidlike and liquidlike states.

\section{Simulation results}

\begin{figure}[h]
\includegraphics[width=1.\linewidth, angle=0]{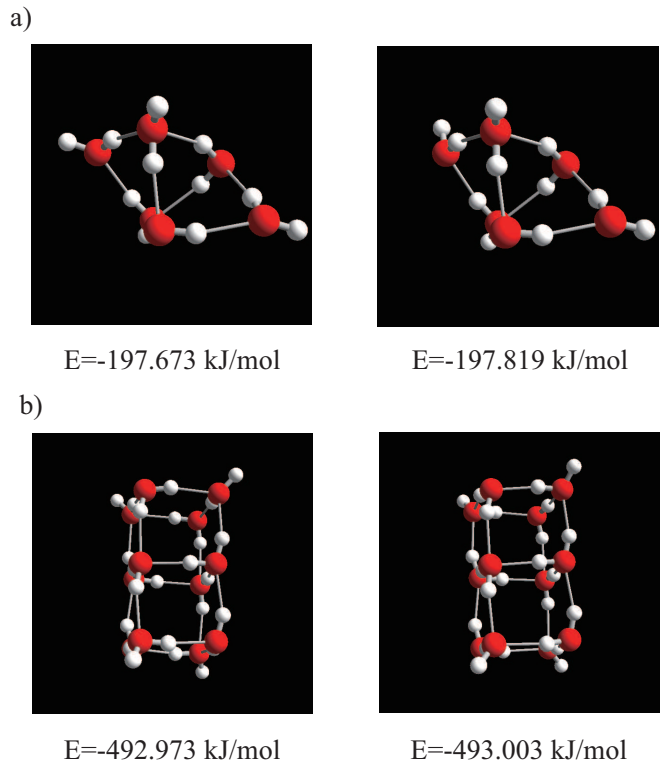}
\caption{ Configurations of water molecules for local potential minima in the solidlike state.
(a) Two different configurations of local potential minima in the solidlike state in $($H${}_2$O$)_{6}$. 
(b) Two different configurations of local potential minima in the solidlike state in $($H${}_2$O$)_{12}$, while 
there are many different local potential minima in it. These configurations are obtained by minimizing 
the potential energy of a water cluster in the solidlike state using the steepest descent method. 
 }
\label{structures}
\end{figure}

Small clusters show rich behaviors, such as hydrogen-bond network rearrangement dynamics \cite{Liu1996},
 size dependence of the melting temperature \cite{Schmidt1998, Kaneko2011}, 
and  dynamical coexistence \cite{Wales1994, Koskinen2007, Nishio2009}.
Here, we study the trapping-time distribution of a solidlike state in small water clusters, 
performing MD simulations of $($H${}_2$O$)_6$ and $($H${}_2$O$)_{12}$. 
Details of MD simulations are described in Appendix A. 
Although it is impossible to define a solid or liquid state for small clusters, 
  some configurations of clusters form an ordered structure and last for a long time, which is reminiscent of a solid state. 
For specific sizes of water clusters, such as eight and twelve, 
some configurations of water clusters last for a long time \cite{Nishio2009, Kaneko2011}. 
Thus, it will be possible to define a solidlike state of small clusters instantaneously  using the potential energy.
 To investigate transition dynamics from solidlike to liquidlike state of water clusters, we propose 
 a definition of solidlike state using  time series of the potential energy. 
 To certain extent, the solidlike state can be defined as the most stable metabasin, which means 
 that the mean escape time from the metabasin is the longest. We have confirmed that configurations of the solidlike state 
 have more ordered structures than those in the liquidlike state (see Figs.~\ref{PEN} and \ref{structures}). 
 Below the transition temperature from liquidlike to solidlike phase, 
 which was obtained in an equilibrium long time simulations,  
 we performed MD simulations of $($H${}_2$O$)_6$ at $T=60$ K 
and $($H${}_2$O$)_{12}$ for $T=135, 138, 142, 149,$ and $155$ K.  
In this temperature region, dynamical coexistence of solidlike and liquidlike states 
is clearly observed.
%{\color{red}Therefore, the solidlike state can be extracted by the potential energy time series.} 

\begin{figure}[h]
\includegraphics[width=.8\linewidth, angle=0]{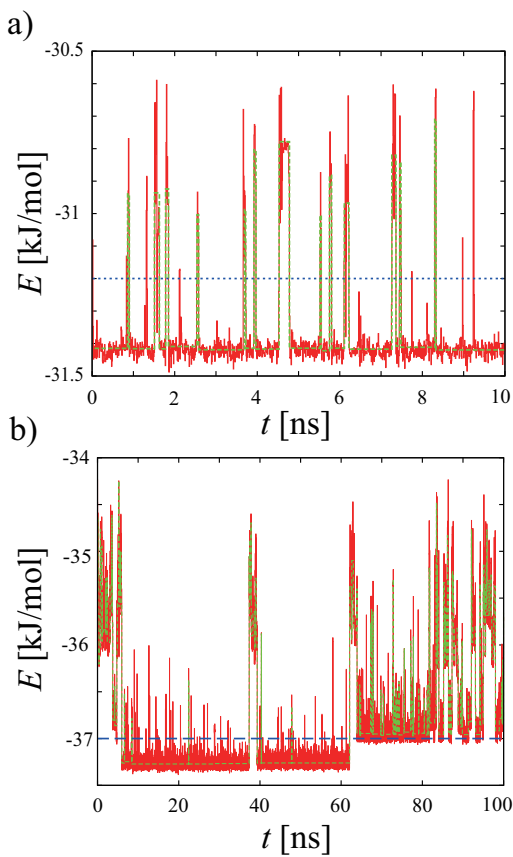}
\caption{ Time series of potential energy $E(t)$ with its coarse-grained one $E_{\rm MA}(t)$ 
for (a) $($H${}_2$O$)_{6}$ at $T=60$ K and (b) $($H${}_2$O$)_{12}$ at $T=135$ K. 
Dashed line represents a boundary between the solidlike and liquidlike state 
 ($E_s=-31.2$ and $-37$ kJ/mol for (a) and (b), respectively).}
\label{PE}
\end{figure}

%{\it Time series of potential energy}.---
As shown in Fig.~\ref{PE}, the potential energy $E(t)$, which is averaged over 10 ps, 
 fluctuates around a constant (green line) for a long time, and the constant changes suddenly, where the 
 green line is a coarse-grained potential energy (see details below). 
Large fluctuations around the constant  imply a change of local potential minimum, % in the phase point, 
while small fluctuations may also imply a change of  local potential minimum. 
On the other hand, a change of the constant corresponds to a change of a metabasin.
It is physically natural to define a solidlike state of small clusters as configurations in a metabasin with 
the longest mean trapping time, which we call the most stable metabasin.

To  search the most stable metabasin, 
 we consider coarse-grained time series of potential energy defined by the followings. 
Time series of potential energy are coarse-grained by a moving average, $E_{\rm MA}(t) \equiv \int_{t_0}^{t} E(t')dt'/(t-t_0)$,
 where a system is in the same metabasin for all $t' \in [t_0,t)$. 
A criterion of whether a system is in the same metabasin is   $|E_{\rm MA}(t) -E(t)| < \delta$, where 
 $\delta$ is a threshold. We note that the threshold $\delta$ should be chosen suitably so as to represent a metabasin.  
 We use $\delta=0.3$ kJ/mol. 
 Even if the criterion does not hold, {\it i.e,} $|E_{\rm MA}(t - \Delta t) -E(t)| \geq \delta$, 
 %{\color{blue} (note that the difference between $E(t)$ and the previous moving average $E_{\rm MA}(t - \Delta t)$ is considered.)}, 
 we consider 
 a system is in the same metabasin if $|E_{\rm MA}(t-\Delta t) -E(t')| < \delta$ for some $t' \in [t,t + 3\Delta t)$, where $\Delta t$ is a minimum time 
 step (10 ps). %This instantaneous fluctuations imply $\beta$-processes.
If the metabasin changes at time $t$, $E_{\rm MA}(t)$ resets, {\it i.e.} $E_{\rm MA}(t)=E(t)$. 
Because metabasins are characterized by the values of $E_{\rm MA}(t)$, $i$th metabasin is defined as $|E_{\rm MA}(t') -E_i| < \epsilon/2$, 
where $\epsilon$ is a parameter characterizing a resolution of metabasin and $E_i$ is an effective energy of $i$th metabasin defined by 
$E_i = E_1 + (i+\frac{1}{2})\epsilon $, where $E_1$ is the lowest energy of $E_{\rm MA}(t')$.
Here, we set $\epsilon=0.1$ kJ/mol. 

Now, we can define a solidlike state as configurations of small clusters in the most stable metabasin, which is obtained by calculating the mean 
trapping times of metabasins. 
We note that our definition of a metabasin is not exactly the same as that in \cite{Buchner2000}, whereas one can successfully obtain 
the most stable metabasin  because instantaneous fluctuations of inherent structures do not affect the trapping times of 
the most stable metabasin.
In fact, we have confirmed our potential energy, which is averaged over 10 ps, represents inherent structures in metabasins with long mean trapping times
 if one neglects instantaneous fluctuations of inherent structures (not shown).
%If the value of $E_{\rm MA}(t)$ is the same, the phase point is in the same metabasin. 
Because the most stable metabasins in $($H${}_2$O$)_{6}$ and $($H${}_2$O$)_{12}$ are the metabasins with 
lowest values of $E_1$,  
we can redefine  configurations with $E_{\rm MA}(t) < E_s$ a solidlike state, 
while we call configurations in the other metabasins a liquidlike state. 
The value of $E_s$ can be defined so as to detect the most stable metabasin. In what follows, we use  
$E_s=-37$ and $-31.2$ kJ/mol for $($H${}_2$O$)_{6}$ and $($H${}_2$O$)_{12}$, respectively. 
For $($H${}_2$O$)_{6}$, as shown in Fig.~\ref{structures}(a), we confirmed that there are at least two different ``cage" structures with lower potential energies,  
which are observed in the disconnectivity graph \cite{Wales2000}. One can see cage structures when $E_{\rm MA}(t)<E_s$ \cite{SM1, SM3}. 
As shown in Fig.~\ref{structures}(b), a solidlike state for $($H${}_2$O$)_{12}$ forms a fused cube structure \cite{Koga2000}, while another state like $E_{\rm MA}(t)= -37$ kJ/mol 
also forms a fused cube structure \cite{SM2, SM3}. 
We note that the value of $E_s$ does not affect dynamics on PENs. %if the above condition is satisfied.

%{\it Geometric distribution in the potential energy network of a water cluster}.---

\begin{figure}
\includegraphics[width=.8\linewidth, angle=0]{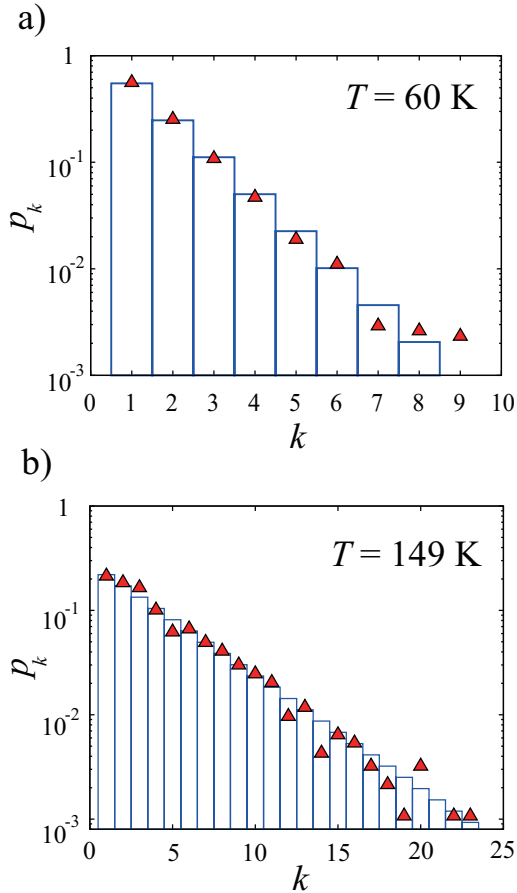}
\caption{ Probability of the number of trials. (a) $($H${}_2$O$)_{6}$ at $T=60$ K. (b) $($H${}_2$O$)_{12}$ at $T=149$ K.
Triangular symbols are the results of the MD simulations. 
Histogram represents the probability of $p(1-p)^{k-1}$ with $p=0.55$ and 0.22 for (a) and (b), respectively.
 }
\label{prob}
\end{figure}

To investigate dynamics on the PEN of the water cluster, we consider 
the probability $p_k$ of the number of large fluctuations of $E(t)$ from $E_{\rm MA}(t)$  until the phase point escapes from a solidlike state. 
The number of large fluctuations 
is defined as the number of events, $|E_{\rm MA}(t) -E(t)| \geq \delta_\beta$, %and $|E_{\rm MA}(t - 2\Delta t) -E(t)| < \delta_\beta$, 
during the solidlike state ($E_{\rm MA}(t)<E_s$).  
The probability $p_k$ depends on $\delta_\beta$ but $\delta_\beta$ does not affect trapping times of the solidlike state. 
Here, we set 
 $\delta_\beta = 0.1$ and 0.3 kJ/mol for $($H${}_2$O$)_{6}$ and $($H${}_2$O$)_{12}$, respectively.  
Surprisingly,  the probability of the number of large fluctuations  for $($H${}_2$O$)_{12}$ cluster 
is described by the geometric distribution, $p_k = p (1-p)^{k-1}$ (see Fig.~\ref{prob}). 
For all temperatures we studied, the geometric distribution appears universally. 
The temperature dependence of $p$ for $($H${}_2$O$)_{12}$ cluster is summarized in Table~\ref{table_water}.  
%{\color{red}Moreover, the relation  (\ref{ave_tau_alpha}), $\langle \tau_\alpha \rangle = \langle \tau_\beta \rangle /p$, holds.}
%(At $T=135$ K, there two different solidlike states. Therefore, we have to 
%consider the probability $p$ for each solidlike state. {\color{blue}But, unfortunately, we do not have enough number of the ensemble 
%to calculate $p_k$ in each solid state.}) 
The result suggests that when the phase point escapes from a local potential minimum, the probability of the escape from the 
 metabasin (solidlike state) is always $p$, which does not depend on the phase point %of local potential minima 
 nor on the number of trials $k$. 
In other words, potential minima within the solidlike state will always be connected to a local potential minimum 
within another metabasin (liquidlike state)  if changes of potential minima imply large fluctuations of the potential energy. 
 %Therefore, dynamics on the PEN of the water cluster can be described by the SHN model. 
This surprising result is one of the main results in this paper. 
We note that if a PEN is not sufficiently connected, the probability 
$p_k$ is not the exponential nor the Poisson distribution because $p_k$ is related to the probability of the first passage time, which is the number of 
steps of the first visit to another metabasin. 
In random walks on lattices with confinements, the distribution of the first passage times is given by a power law 
with a cutoff \cite{Redner2001}. If a network within a metabasin is not sufficiently connected, a random walk in the network is similar to that on a lattice 
with a confinement \cite{Redner2001}. However, 
the probability $p_k$ does not show a power law even in short-time steps. 
%Therefore, the PEN of the water cluster will have a small world character like a scale-free network \cite{Doye2002}.

%Figure 4

\begin{figure}
\includegraphics[width=.8\linewidth, angle=0]{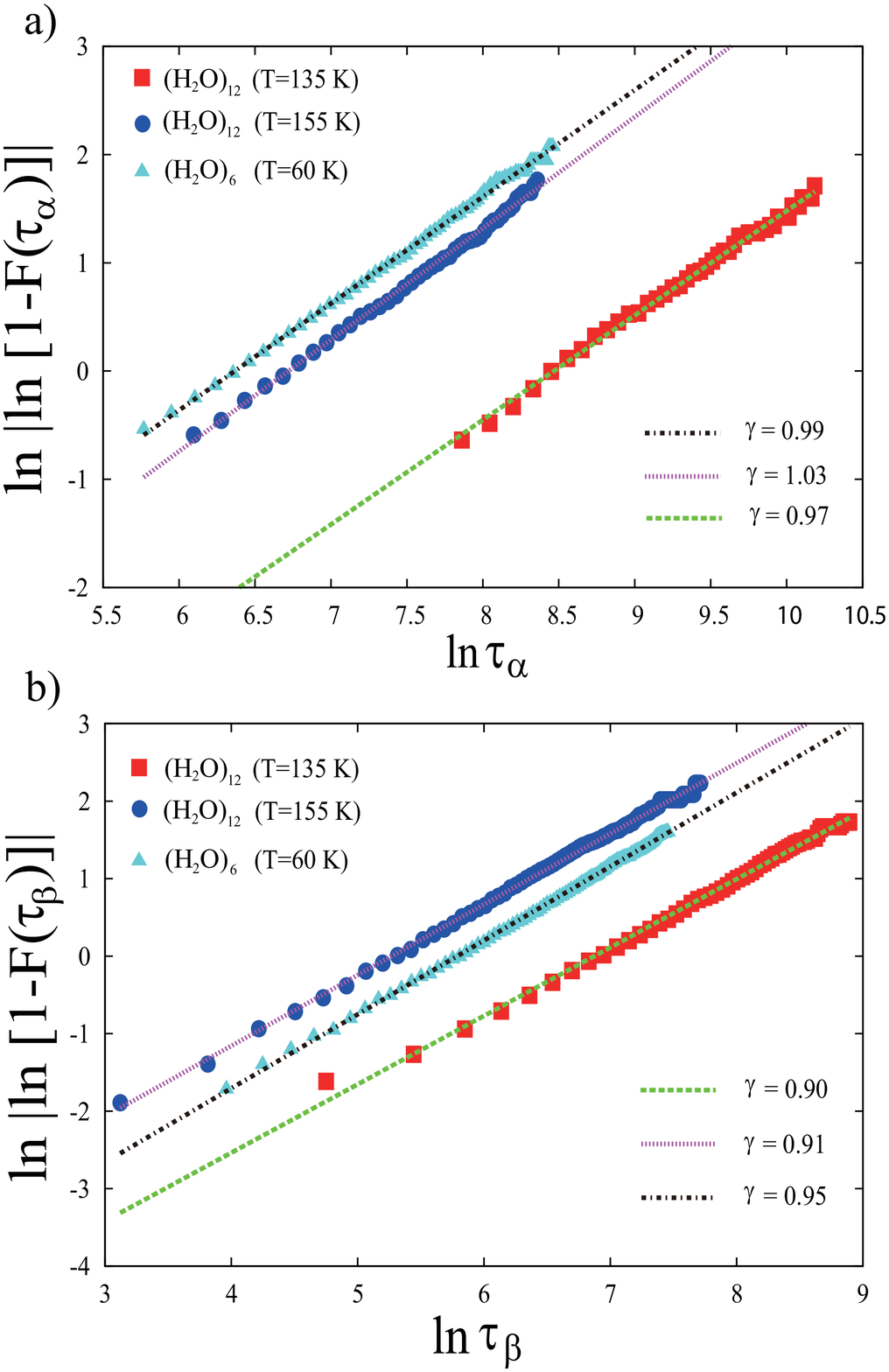}
\caption{ Weibull plot, {\it i.e.}, (a) $\ln | \ln [1-F(\tau_{\alpha})]|$ vs. $\ln \tau_\alpha$ and (b) 
$\ln | \ln [1-F(\tau_{\beta})]|$ vs. $\ln \tau_\beta$. Symbols are the 
results of the MD simulations. Lines are the fitting lines. 
The slopes of the fitting lines indicate the exponent $\gamma$ of
 the Weibull distribution (\ref{weibull}). The Weibull exponents for the trapping-time distribution of $\tau_\alpha$ are almost $\gamma=1$, 
 which implies the exponential distribution. 
}
\label{fig_weibull_plot}
\end{figure}

%{\it Trapping-time distributions corresponding to $\alpha$ and $\beta$ processes}.---
Here, we consider trapping times 
$\tau_\alpha$ in the solidlike state and 
%which is defined as $E_{\rm MA}(t) < E_s$. %We set $E_1=-36.5$ because clusters form a solidlike state in this region. 
 interevent times $\tau_\beta$ of large fluctuations, {\it i.e.},
  $|E_{\rm MA}(t) -E(t)| \geq \delta_\beta$ 
%$|E_{\rm MA}(t - 2\Delta t) -E(t)| < \delta_\beta$ 
and $E_{\rm MA}(t) < E_s$. We find that 
the distributions of $\tau_\beta$  follow the Weibull distribution (see Fig.~\ref{fig_weibull_plot}b):
\begin{equation}
\int_0^{\tau_\beta} P(\tau)d\tau= 1 - \exp [- (\tau_\beta/b)^{\gamma}],
\label{weibull}
\end{equation}
where $b$ is the relaxation time for the Weibull distribution. 
The Weibull exponents $\gamma$ obtained by the Weibull plot (Fig.~\ref{fig_weibull_plot}) and the mean trapping time $\langle \tau_\beta\rangle$   
are summarized in Table~\ref{table_water}. 
%The exponents $\gamma$ are equal to around 0.90, 0.92, and 0.91 for $T=$ 135, 149, and 155 K, respectively (see Fig.~\ref{dist}). 
On the other hand, as expected in the SHN model, 
the distributions of $\tau_\alpha$ follow the exponential distribution:
\begin{equation}
F(\tau_\alpha) = 1 - e^{- \tau_\alpha/a},
\label{exp}
\end{equation}
where $a$ is the relaxation time. 
The mean trapping time $\langle \tau_\alpha \rangle$ and the relaxation time $a$ 
are summarized in Table~\ref{table_water}. 
Using the probability $p$ and the mean trapping times  $\langle \tau_\beta \rangle$, we can estimate the mean $\langle \tau_\alpha \rangle$ and 
the relaxation times $a$ by $\langle \tau_\alpha \rangle= a = \langle \tau_\beta \rangle/p$
[see Eqs.~(\ref{ave_tau_alpha}) and (\ref{exponential})]. 
The estimated values $\langle \tau_\beta \rangle /p$ given in Table~\ref{table_water}
 are consistent with $\langle \tau_{\alpha} \rangle$ as well as the relaxation time $a$ obtained by the exponential fittings of $P(\tau_\alpha)$.   
  Therefore,  the relations $\langle \tau_\alpha\rangle = \langle \tau_\beta \rangle /p$ and $a=\langle \tau_\alpha \rangle$ 
  are valid in water clusters.

Although the Weibull exponent of $P(\tau_\beta)$ for $($H${}_2$O$)_{6}$ at $T=60$ K is almost unity, the result of the exponential distribution 
for $P(\tau_\alpha)$ is not trivial. In the disconnectivity graph for TIP4P $($H${}_2$O$)_{6}$,  
 there are two different potential minima in a solidlike state, defined by $E_{\rm MA}(t) < 31.2$ kJ/mol \cite{Wales2000}.  
 Because the exponential distribution cannot appear for the trapping-time distribution within many potential minima if the connectivity 
is sparse and inhomogeneous, the exponential 
distribution in  $P(\tau_\alpha)$ originates from homogeneous connectivity between nodes within a solidlike state 
and those within a liquidlike state.  In other words, escape probability from a metabasin (solidlike state) is always constant and does not depend 
on a node (potential minimum). 
We note that values of potential energy in the disconnectivity graph  are not directly connected 
to our potential energy  because our potential energy is averaged value. 

%Figure 5
    
\begin{table}[h]%The best place to locate the table environment is directly after its first reference in text
\caption{%\label{tab:table1}%
The probability $p$, the Weibull exponent $\gamma$, the mean trapping time $\langle \tau_\beta \rangle$, $\langle \tau_\beta \rangle/p$,
the mean trapping time $\langle \tau_\alpha \rangle$, and the relaxation time $a$ for $($H${}_2$O$)_{12}$.
}
\begin{ruledtabular}
\begin{tabular}{ccccccc}
%\hline
\textrm{$T$ [K]}&
\textrm{$p$}&
\textrm{$\gamma$}&
%\multicolumn{1}{c}{\textrm{Decimal}}&
\textrm{$\langle \tau_\beta \rangle$ [ns]}&
\textrm{$\langle \tau_\beta \rangle/p$ [ns]}&
\textrm{$\langle \tau_\alpha \rangle$ [ns]}&
\textrm{$a$ [ns]}\\ %$\langle \tau_\alpha \rangle$ \footnote{obtained by the exponential fitting.}}\\
\colrule
%\hline
135 & 0.22 & 0.90 & 1.15  & 5.25 & 5.19 & 5.39\\
138 & 0.22 & 0.87 & 0.96  & 4.36 & 4.12 & 4.13\\
142 & 0.22 & 0.88 & 0.65  & 2.95 & 2.84 & 2.89\\
149 & 0.22 & 0.92 & 0.37  & 1.68 & 1.50 & 1.48\\
155 & 0.25 & 0.91 & 0.23  & 0.92 & 0.89 & 0.84\\
%\hline
\end{tabular}
\end{ruledtabular}
\label{table_water}
\end{table}

\section{ Discussion} 

\subsection{Generalization of the simple homogeneous network model}

We have found that transitions between solidlike and liquidlike state  of water clusters are well described by 
the SHN model. The exponential distribution of 
trapping times within the solidlike state is universal in such a PEN even when the trapping-time distribution for local potential 
minima is not exponential. We note that trapping times in the solidlike state considered here 
are different from hopping times from metabasins \cite{Denny2003}, which are distributed according to a non-exponential distribution 
(we consider trapping times of only one specific metabasin). 
Non-exponential trapping-time distributions such as power laws are observed in Hamiltonian system \cite{Miyaguchi2007a}, 
supercooled liquids,  
systems close to the glass transition \cite{Debenedetti2001,Doliwa2003,Doliwa2003a}, and biological 
systems \cite{Weigel2011, Akimoto2011}. One of the well-known mechanisms of a non-exponential distribution is 
a random energy barrier model.  If the heights of barriers are distributed according to the exponential distribution, 
the distribution of trapping times that a particle or the phase point is confined by the barriers follows a power law because of the Arrhenius law. 

\begin{figure}
\includegraphics[width=1.\linewidth, angle=0]{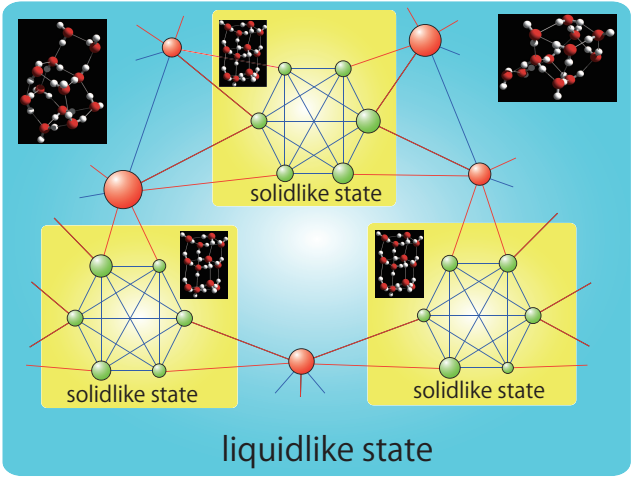}
\caption{ Schematic picture of a generalized SHN model. Potential minima within solidlike states (green spheres) 
are not directly connected to those within solidlike states.  
}
\label{energy_landscape_non-exp}
\end{figure}

Here, we generalize the SHN model to provide another origin of a non-exponential distribution in a solidlike state. 
In the SHN model, there exists only one metabasin in the solidlike state. However, there are many metabasins 
in a solidlike state for large systems. On the basis of the SHN model, we consider transition dynamics between 
specific metabasins and the other metabasins. We assume that there are $n$ metabasins in a solidlike state and that these metabasins 
are not directly connected with each other (see Fig.~\ref{energy_landscape_non-exp}). We also assume that escape dynamics from a metabasin within the solidlike state can be described 
by the SHN model, {\it i.e.}, the exponential distribution. 
In particular, the trapping-time distribution in the $i$th metabasin is described by the exponential 
distribution with a relaxation time $\tau_i$. With the aid of the disconnectivity of the metabasins within the solidlike 
state, the trapping-time distribution within the solidlike state can be described by 
\begin{equation}
P(\tau) = \sum_{i=1}^n p_i \tau_i^{-1} \exp (-\tau/\tau_i),
\label{superposition}
\end{equation}
where $p_i$ is the probability of a transition from the liquidlike state to the $i$th metabasin within the solidlike state. 
Here we assume that the probability $p_i$ does not depend on a node in the liquidlike state. 
 Therefore, the trapping-time distribution is a non-exponential distribution 
if the relaxation times  are widely distributed. In particular, if the inverse of the relaxation times $\nu$ 
are distributed according to the Weibull distribution $\rho(\nu)=(\eta+1)\nu^\eta e^{-\nu^{\eta+1}}$, the trapping-time distribution $P(\tau)$ is given by a power law:
\begin{eqnarray}
%P(\tau) = \int_0^\infty  \nu^{\eta+1} e^{-\nu^{\eta+1}} e^{-\nu \tau} d\nu \sim (\eta +1) \tau^{-(2+\eta)} 
P(\tau) = \int_0^\infty  \rho(\nu) \nu e^{-\nu \tau} d\nu \sim (\eta +1)^2 \tau^{-(2+\eta)} 
\end{eqnarray}
for $\tau \rightarrow \infty$. 
%We have provided a novel route to a non-exponential trapping time distribution in the solidlike state. 
%Finally, we have found the trapping-time distribution within a solid state obeys a superposition of exponential distributions if 
%there are different metabasins in a solidlike state. 
Well-known origin of a power-law trapping-time distribution are
 a random energy landscape and inner degrees of freedom \cite{bouchaud90}. 
The above scenario provides another route to a power-law trapping-time distribution, {\it i.e.}, a superposition of exponential distributions, 
which is basically originated from the connectivities in the PEN. We note that this scenario is completely different from heterogeneous scenario 
in supercooled liquids \cite{Richert2002} 
because the heterogeneous scenario provides a non-exponential trapping-time distribution of a particle, whereas our scenario provides 
that of a solidlike state.
%(a superposition of exponential distributions with different relaxation times becomes a power-law distribution 
%if the relaxation times are distributed according to the Weibull distribution.).
%It would be interesting if one could see a power-law trapping-time distribution in clusters.
%These results are consistent with the simple homogeneous network model we proposed here. 

\begin{figure}
\includegraphics[width=1.\linewidth, angle=0]{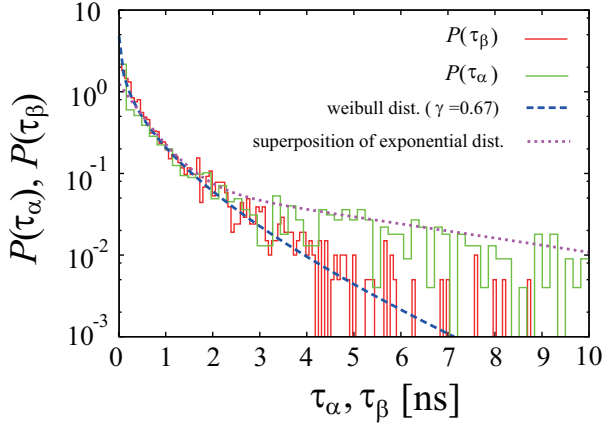}
\caption{ Probability density functions for $\tau_{\alpha}$ and $\tau_\beta$ in a semi-log scale for $T=135$ K. 
Histograms are the results of the MD simulations. The value of $E_s$ is set to be $-36.5$ kJ/mol.
The dashed line is a fitting curve of the Weibull distribution (\ref{weibull}) obtained by the Weibull plot. 
%The Weibull exponents at $T=135, 149$, and 155 K are 0.90, 0.92, and 0.91, respectively. 
The dotted line is a fitting curve of a superposition of the exponential distribution (\ref{superposition}) for $n=2$.
The fitting parameters are $p_1=0.4$ and $p_2=0.6$. The two relaxation times are obtained by the mean 
trapping times for $E_{\rm MA}<-37$ kJ/mol and $-37<E_{\rm MA} <-36.5$ kJ/mol ($\tau_1=5$ ns and $\tau_2=0.5$ ns). 
}
\label{dist_superposition}
\end{figure}

Figure~\ref{dist_superposition} shows the trapping-time distributions of $\tau_\alpha$ and $\tau_\beta$, 
where $E_s$ is set to $-36.5$ kJ/mol, below which there are two metabasins (fused cube structure). 
A superposition of the exponential distributions, Eq.~(\ref{superposition}), is in good agreement with 
numerical simulations $(n=2)$, indicating that two metabasins are not directly connected each other.
%We note that a non-exponential distribution does not appear if all relaxation times are the same.

%Figure 6

%Figure 7

\subsection{Origin of the Weibull distribution in the trapping-time distribution}

%It is reasonably expected that scale-free networks are compatible 
%with our results in a sense that all potential minima are connected within a few paths through a hub 
%in scale-free networks\cite{Barabasi1999, Hwang2012}.
%A simple homogeneous model will be valid for other clusters.

We have also found that interevent times of large fluctuations in the potential energy within the solidlike state
 are distributed according to the Weibull distribution. 
It has been known that there are two mechanisms generating the Weibull distribution in the trapping-time distribution. One is a random energy barrier. 
If the random energy barriers are distributed according to the double exponential, the trapping-time distribution obeys the 
Weibull distribution. For example, when the distribution of barrier heights $\Delta E >0$ is given by 
\begin{equation}
\Pr ( \Delta E <x) = 1 - \exp(1 - e^{E_0x}), 
\label{double_exp}
\end{equation}
the distribution of the trapping time $\tau_\beta (>\tau_0)$ can be described by the Arrhenius law:
\begin{eqnarray}
\Pr (\tau_\beta<x) &=& \Pr ( \tau_0 \exp (\Delta E/k_BT) <x) 
%\\
%&=& \Pr \{ \Delta E \leq  k_BT \ln (x/\tau_0)\}
%\\
%&=& 1 - \exp (1 - e^{  E_0^{-1}\ln (x/\tau_0)^{k_BT}} )
\\
&=& 1-\exp ( 1 - (x/\tau_0)^{k_BT/E_0}).
\label{Weibull_REL}
\end{eqnarray}
Therefore, the Weibull exponent depends linearly on the temperature. 
The mean of $\tau_\beta$ is given by $\langle \tau_\beta \rangle = \tau_0 \Gamma (1+E_0/k_BT, 1)$, 
where $\Gamma(s,x)=\int_x^\infty  t^{s+1}e^{-t}dt$. 

The other mechanism of the Weibull distribution is correlated time series. If time series are 
strongly correlated like a $1/f^{\beta}$ spectrum, 
recurrence times to exceed a threshold are distributed according to the Weibull distribution \cite{Altmann2005}. 
If potential energy time series within the metabasin are strongly correlated, the interevent-time distribution of $E(t)$ of 
large fluctuations from $E_{\rm MA}(t)$
 follows the Weibull distribution. 

\begin{figure}
\includegraphics[width=1.\linewidth, angle=0]{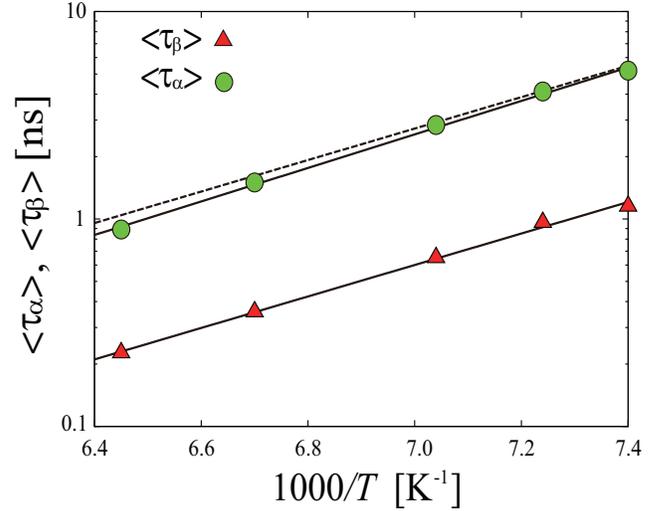}
\caption{ The mean trapping times $\langle \tau_\alpha \rangle$ and $\langle \tau_\beta \rangle$ vs. $1000/T$ in  $($H${}_2$O$)_{12}$ 
water cluster.
 The values of $E_s$ are set to be $-37, -37, -36.9, -36.5,$ and  $-36.5$ kJ/mol for 
$T=135, 138, 142, 149,$ and 155 K, respectively. 
Circles and triangles  are the results for 
$\langle \tau_\alpha \rangle$ and $\langle \tau_\beta \rangle$, respectively. Solid lines are the linear fitting lines: 
$\langle \tau_\alpha \rangle = \tau_0 \exp( 1000A/T)$ and $\langle \tau_\beta \rangle = \tau_1 \exp( 1000B/T)$ 
($\tau_0=5.6\times 10^{-6}$, $A=1.86$, $\tau_1=2.9\times 10^{-6}$, and $B=1.74$). 
Dashed line represents $\tau_0 \exp(  1000B/T)/p$.
}
\label{arrhenius}
\end{figure}

  Arrhenius plot for the mean trapping times $\langle \tau_\beta \rangle$ and $\langle \tau_\alpha \rangle$ is shown in 
  Fig.~\ref{arrhenius}. %where the values of $E_s$ are set to be $-37, -37, -36.9, -36.5,$ and  $-36.5$ kJ/mol for 
%$T=135, 138, 142, 149,$ and 155 K, respectively. 
  Arrhenius law, $\langle \tau \rangle = \tau_0 \exp(\Delta E/k_B T)$, seems to hold for both $\langle \tau_\beta \rangle$ 
  and $\langle \tau_\alpha \rangle$, whereas the distribution of $\tau_\beta$ is not exponential and there are local potential minima in 
 the solidlike state.  
 % In other words, escapes from {\color{red}the solidlike state} can be characterized by a single effective potential, 
 % $\langle \tau_\alpha \rangle = \tau_0' \exp(\Delta E'/k_B T)$. 
  These results are not consistent with the random energy barrier scenario because the mean interevent time $\langle \tau_\beta \rangle$ 
  does not satisfy the relation $\langle \tau_\beta \rangle \propto \exp(\Delta E_\beta/k_B T)$ but satisfy $\langle \tau_\beta \rangle 
  \propto \Gamma (1+ E_0/k_BT, 1)$ in the random energy barrier model. 
  Moreover, it is not clear whether the Weibull exponents given in Table I depends linearly on temperature as in Eq.~(15).
  %\ref{Weibull_REL}. 
  Therefore, the origin of the Weibull distribution is still controversial.
%  We note that the relation $\langle \tau_\alpha \rangle = \langle \tau_\beta \rangle/p$ can be generalized as
 % $\langle \tau_\alpha (T) \rangle = \langle \tau_\beta (T) \rangle/p(T)$. 
It is interesting to clarify the origin of the Weibull distribution in small clusters. 
This problem is left for a future work. 

%Figure 8

%Finally, we assume that there is only one metabasin in a solidlike state for simplicity. Generally, there are many metabasins in a solidlike state. 
%Therefore, a generalization of a simple homogeneous model is desired, which is left for a future work. 
%This generalization could effect non-exponential trapping-time distributions (slow relaxation phenomena). 

\section{Conclusion}

Introducing a concept of a solidlike state in small clusters, we have shown that 
dynamics regarding transitions between solidlike and liquidlike states in water clusters 
are well described by the SHN model proposed here, where  the mean trapping time $\langle \tau_\alpha \rangle$
of a solidlike state is given by the mean interevent time $\langle \tau_\beta \rangle$ of large fluctuations of $E(t)$ within the 
solidlike state through the relation $\langle \tau_\alpha \rangle = \langle \tau_\beta \rangle/p$. 
Unlike supercooled liquids, 
the trapping-time distribution of a solidlike state, which is related to $\alpha$-process, follows the exponential distribution. 
Thus, $\alpha$ processes in water clusters are completely different from those in supercooled liquids and 
glass transition. 
The exponential distribution is originated from 
the homogeneous connectivity between local potential minima within the metabasin and those within the other metabasins.

\appendix

\section{Molecular dynamics simulations}

The TIP4P water model \cite{Jorgensen1983} is used in the simulations of the water cluster. 
Initial coordinates are taken from the global minima of the cluster 
at 0 K \cite{Wales1998}, which are available from the Cambridge Cluster Database \cite{Wales}.
 Initially, the momentum and the angular momentum are set to be zero. 
The MD simulations are performed in conventional canonical ensembles. 
The temperatures are controlled by Nos\'{e}-Hoover thermostat.
The mass of thermostat per molecule is chosen to $0.1$ ps${}^2$ kJ/mol
  so that the frequency of thermostat variable becomes same as that of the vibration of oxygen
  to avoid the artificial dynamics by the thermostat. 
  We confirmed that the mass of thermostat did not affect our results. 
%and fixed at $T= 120, 135,149,$ and 155 K, at which a water cluster is in a solidlike phase.
 The simulations were performed in free boundary condition. 
  The integration scheme is the velocity Velret algorithm with SHAKE/RATTLE method, in which time interval is 0.5 fs.
We used the MD simulation data of 250 ns for different ten initial conditions, in which the initial 0.05 ns was excluded for 
an equilibration.

\bibliography{water}

%%%%%%%%%%%%%%%%%%%%%%%%%%%%%%%%%%%%%%%%%%%%%%%%%%%%%%%%%%%%%%%%%%%%%
%% The same is true for Supporting Information, which should use the
%% suppinfo environment.
%%%%%%%%%%%%%%%%%%%%%%%%%%%%%%%%%%%%%%%%%%%%%%%%%%%%%%%%%%%%%%%%%%%%%
%\begin{suppinfo}

%This will usually read something like: ``Experimental procedures and
%characterization data for all new compounds. The class will
%automatically add a sentence pointing to the information on-line:

%\end{suppinfo}

%%%%%%%%%%%%%%%%%%%%%%%%%%%%%%%%%%%%%%%%%%%%%%%%%%%%%%%%%%%%%%%%%%%%%
%% The appropriate \bibliography command should be placed here.
%% Notice that the class file automatically sets \bibliographystyle
%% and also names the section correctly.
%%%%%%%%%%%%%%%%%%%%%%%%%%%%%%%%%%%%%%%%%%%%%%%%%%%%%%%%%%%%%%%%%%%%%
%\bibliography{achemso-demo}

%%%%%%%%%%%%%%%%%%%%%%%%%%%%%%%%%%%%%%%%%%%%%%%%%%%%%%%%%%%%%%%%%%%%%
%% The "tocentry" environment can be used to create an entry for the
%% graphical table of contents.
%%%%%%%%%%%%%%%%%%%%%%%%%%%%%%%%%%%%%%%%%%%%%%%%%%%%%%%%%%%%%%%%%%%%%

\end{document}